\documentclass[12pt]{article}
\usepackage{epsfig}
\usepackage{graphicx}
\usepackage{lineno}
\usepackage{amsmath}
\usepackage{hyperref}
\usepackage{float}

\def\beq{\begin{equation}}
\def\eeq#1{\label{#1}\end{equation}}
\def\eeqn{\end{equation}}

\def\beqa{\begin{eqnarray}}
\def\eeqa#1{\label{#1}\end{eqnarray}}
\def\eeqan{\end{eqnarray}}

\let\bar=\overbar

\def\Dslash{\not{\hbox{\kern-4pt $D$}}}
\def\dslash{\not{\hbox{\kern-2pt $\del$}}}

\def\msb{{\bar{\ssstyle M \kern -1pt S}}}

\def\Title#1{\begin{center} {\Large {\bf #1} } \end{center}}

\begin{document}

\Title{Lepton-flavour universality tests with semi-leptonic B decays at LHCb}

\bigskip\bigskip


\begin{raggedright}

{\it Benedetto Gianluca Siddi\index{Siddi, B.G.} on behalf of LHCb collaboration\footnote{Talk presented at the APS Division of Particles and Fields Meeting (DPF 2017), July 31-August 4, 2017, Fermilab. C170731}\\
Dipartimento di Fisica\\
Universit\`a degli studi di Ferrara\\
42122 Ferrara, ITALY\\
CERN, CH-1211 Geneva 23, SWITZERLAND}
\bigskip\bigskip
\end{raggedright}

\section{Introduction}

Lepton universality, a symmetry of the Standard Model (SM), predicts equal coupling between gauge bosons and the three lepton families. Standard Model extensions give additional interactions, implying in some cases a stronger coupling with the third generation of leptons. Semileptonic decays of b hadrons (hadrons containing a b quark) provide a sensitive probe to such new physics effects. The presence of additional charged Higgs bosons or leptoquarks \cite{Tanaka1995,Dorsner20161}, required by such SM extensions, can have significant effect on the semileptonic decay rate of $B^0\rightarrow D^{*-} \tau^+ \nu_\tau$ . Since uncertainties due to hadronic effects cancel to a large extent, the SM prediction for the ratios between branching fractions of semitauonic decays of \textit{B} mesons relative to decays involving lighter lepton families, such as
$R(D^{*+}) = \frac{\mathcal{B}(\bar{B}^0 \rightarrow D^{*+} \tau^- \bar{\nu}_\tau)}
{\mathcal{B}(\bar{B}^0 \rightarrow D^{*+} \mu^- \bar{\nu}_\mu)}$
and $R(D^+) = \frac{\mathcal{B}(\bar{B}^0 \rightarrow D^{+} \tau^- \bar{\nu}_\tau)}{\mathcal{B}(\bar{B}^0 \rightarrow D^{+} \mu^- \bar{\nu}_\mu)}$, are known with an uncertainty at the percent level \cite{Bigi:2016mdz,Fajfer:2012vx}.
The combination of experimental measurements performed by BaBar\cite{Lees:2013uzd,Lees:2012xj}, Belle \cite{Huschle:2015rga,Sato:2016svk} and LHCb \cite{Aaij:2015yra} observing the channel where the $\tau$ decays in leptons or in one prong particle \cite{PhysRevLett.118.211801}, gives a deviation from the SM prediction of about 4 $\sigma$. It is therefore important to perform additional measurements in this sector in order to improve the precision and confirm or disprove this deviation. Another possibility is to perform these measurements using $\tau^- \to \pi^-\pi^+\pi^-\nu_{\tau}$ decay channel (3-prong decay). A measurement of $R(D^{*-})$ using 3-prong $\tau$ decays performed by LHCb is reported. This report, references the article \cite{Aaij:2017uff}, submitted to arXiv.

%
%

\section{Semitauonic decays of b hadrons}
	Semitauonic decays of $B^0$ and $B^+$ mesons are successfully studied in \textit{B-factories} with high-statistics samples. Despite the hadronic enviroment, LHCb is also able to study such kind of decays and extend to other \textit{b} hadrons ($B_s^0$, $\Lambda_b^0$, $B_c^+$) thanks to the high boost and the excellent vertexing capabilities.
	However, there are some analysis challenges in a hadronic enviroment. Rest frame variables are needed in order to distinguish between signal and background. Backgrounds due to unreconstructed charged and neutral particles must be suppressed. A normalization channel, necessary to reduce at minimum systematic uncertainties, must be identified.

  A way to measure $R(D^{*-})$ is to reconstruct the $\tau$ in the 3-pions decay channel. This is a semileptonic decay without charged leptons in the final state. In this case, the measured quantity is:
\begin{equation}
  K(D^{*-}) = \frac{\mathcal{B}(B^0 \rightarrow D^{*-} \tau^+ \nu_\tau)}{\mathcal{B}(B^0 \rightarrow D^{*-} \pi^+\pi^-\pi^+)} = \frac{N_{sig}}{N_{norm}}\frac{\epsilon_{norm}}{\epsilon_{sig}}\frac{1}{\mathcal{B}(\tau^+\rightarrow 3\pi(\pi^0)\bar{\nu}_\tau)}
\end{equation}

\noindent where $N_{sig}$ and $N_{norm}$ are the signal and the normalization yields, $\varepsilon_{\mathrm{sig}}$ and
$\varepsilon_{\mathrm{norm}}$ are the efficiencies for the signal and
normalization decay modes, respectively.

The two decays modes share the same visible final state and most of the systematic uncertainties will cancel. Then, $R(D^{*-})$ can be measured using:
\begin{equation}
  R(D^{*-}) = K(D^{*-}) \times \frac{\mathcal{B}(B^0 \rightarrow D^{*-} 3\pi)}{\mathcal{B}(B^0 \rightarrow D^{*-} \mu^+ \nu_\mu)}
\end{equation}

The quantities multiplying $K(D^{*-})$ are taken as external inputs \cite{PDG2017}.

\section{Events selection and main backgrounds}

\subsection{Detached-vertex method}
The most abundant background source is due to hadronic \textit{B} decays into $D^{*-} 3\pi X$, where $X$ denotes unreconstructed particles.
Such background has a branching fraction that is about 100 times that of the signal.
\begin{align*}
  \frac{\mathcal{B}(B^0 \rightarrow D^* 3 \pi X)}{\mathcal{B}(B^0 \rightarrow D^{*-} \tau^+ \nu)_{SM}} \sim 100
\end{align*}

\begin{figure}[ht]
  \centering
  \includegraphics[width=0.6\textwidth]{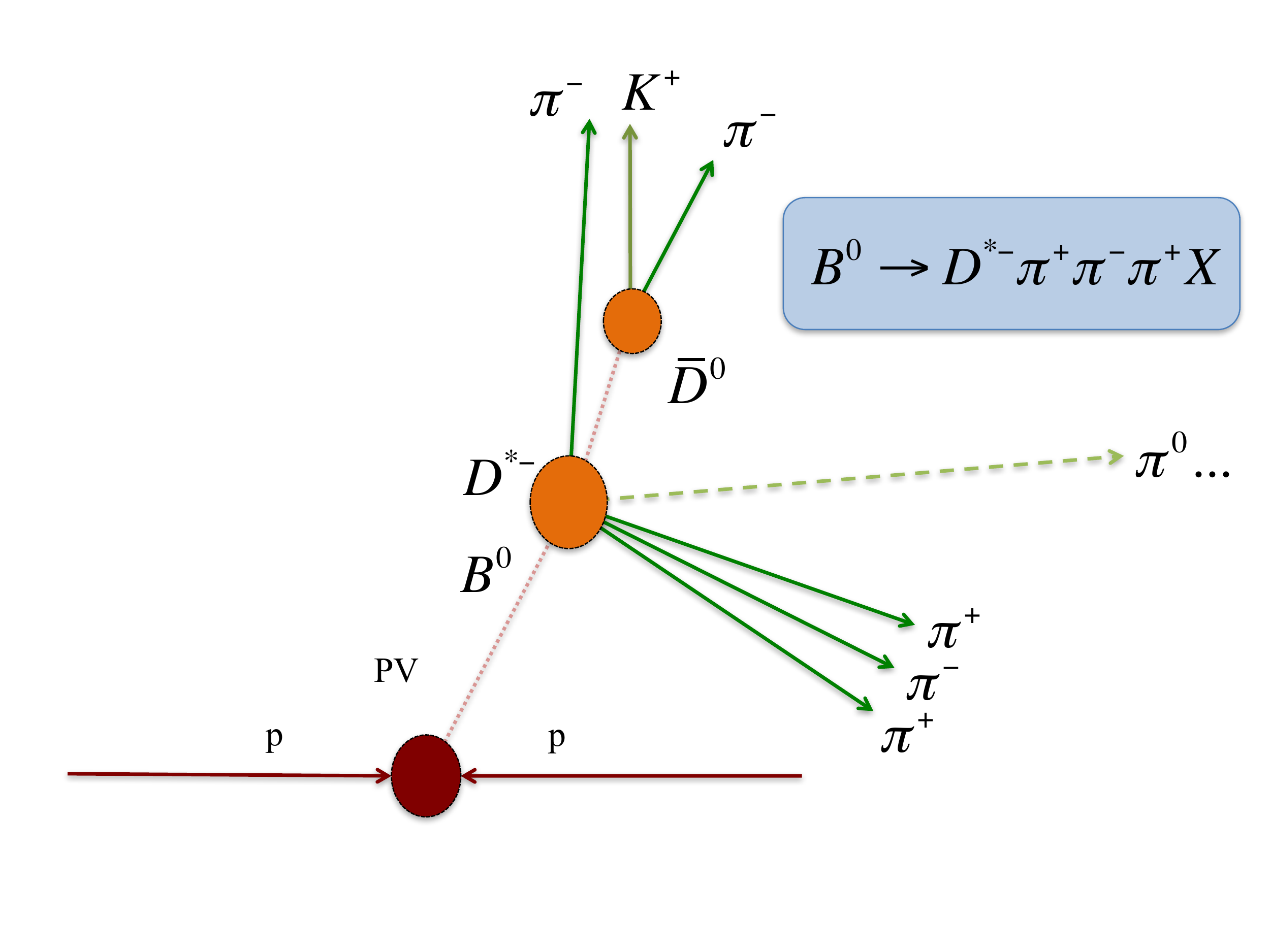}
  \caption{Schematic view of the $B^0 \rightarrow D^{*-} 3\pi X$ decay in which the 3-pions comes from the $B^0$ vertex.}
  \label{fig:vertex}
\end{figure}

In Fig.~\ref{fig:vertex} a schematic view of the $B^0 \rightarrow D^{*-} 3\pi X$ decay is shown.

\begin{figure}[ht]
  \centering
    \includegraphics[width=0.6\textwidth]{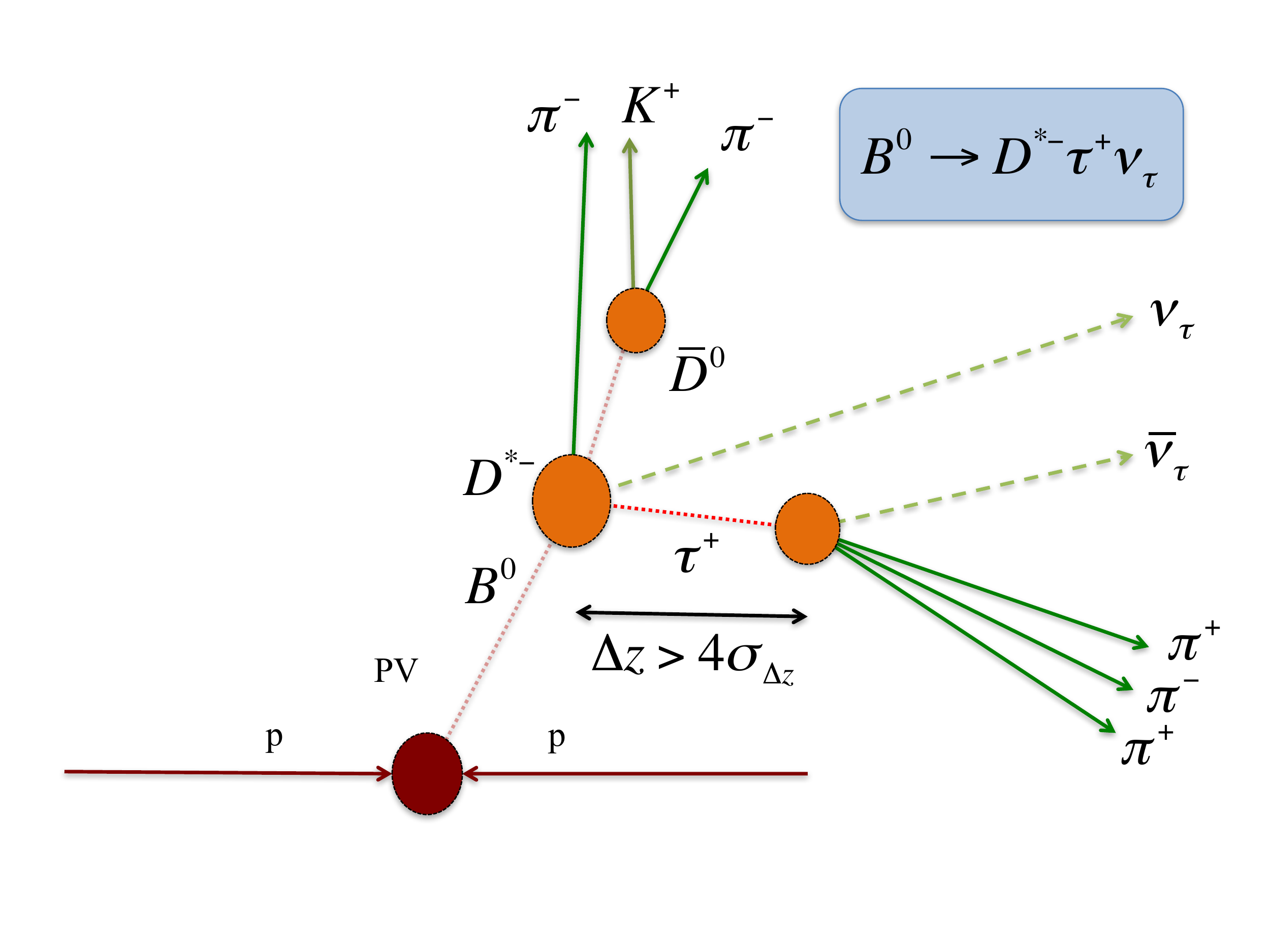}
    \caption{Schematic view of the detached-vertex requirement.}
    \label{fig:inv_vertex}
\end{figure}

In the signal case, because of the significant $\tau$ lifetime and boost along the forward direction, the $3\pi$ system is detached from the $B^0$ vertex.
Thanks to the excellent vertex resolution, it is possible to require a detached vertex topology for the $\tau$ particle. This consists on the requirement that the $\tau$ vertex is downstream with respect to the $B^0$ vertex along the beam direction ($\Delta_z$) with a significance of at least $4\sigma$. A schematic view of this requirement is shown in Fig.~\ref{fig:inv_vertex}. Using this requirement, the background due to $B\rightarrow D^{*-}3\pi X$ is suppressed by about 3 orders of magnitude.

\begin{figure}[ht]
  \centering
  \includegraphics[scale=0.6]{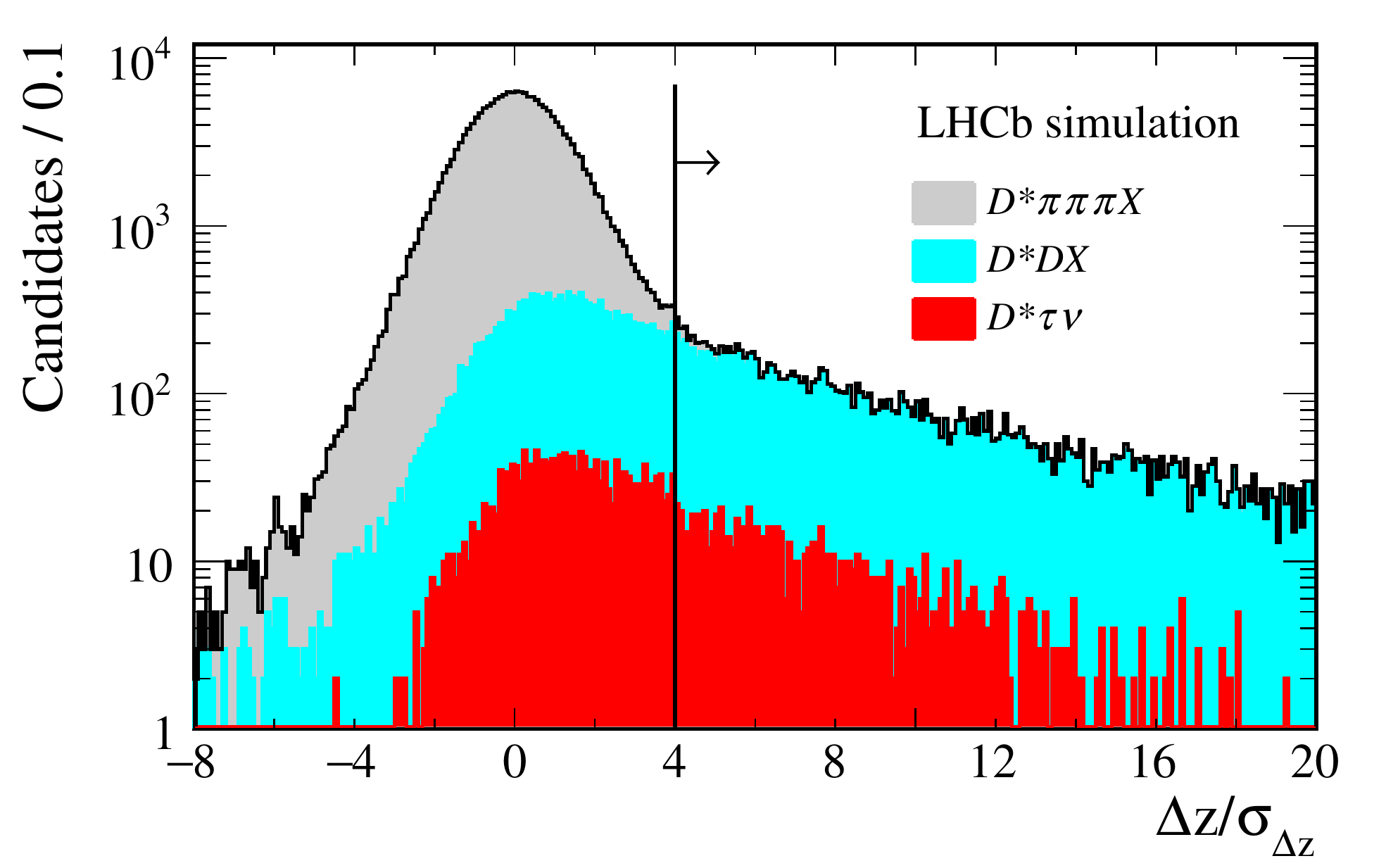}
  \caption{Distribution of the vertex significance ($\Delta_z / \sigma$) for different background sources.}
  \label{fig:inv_vertex_2}
\end{figure}

The distribution of the quantity $\Delta_z / \sigma_{\Delta_z}$ for different background sources is presented in Fig.~\ref{fig:inv_vertex_2}. The contribution from $B\to D^{*-}3\pi X$ decays (grey) is suppressed, after a cut of $\Delta_z / \sigma_{\Delta_z} > 4$.

\subsection{Double charm decays}
After the detached-vertex requirement, the remaining background consists of $B$ hadron decays where the $3\pi$ vertex is transported away from the $B$ vertex by a charm carrier, such as $D^0$, $D^+$ and $D^+_s$ mesons. This background has a yield that is about 10 times the signal.
\begin{align*}
  \frac{\mathcal{B}(B^0 \rightarrow D^* D_{(s)}^{(*)}; D_{(s)}^{(*)} \rightarrow 3\pi +X)}{\mathcal{B}(B^0\rightarrow D^* \tau \nu)}\sim 10
\end{align*}

\begin{figure}[ht]
  \centering
  \includegraphics[scale=0.5]{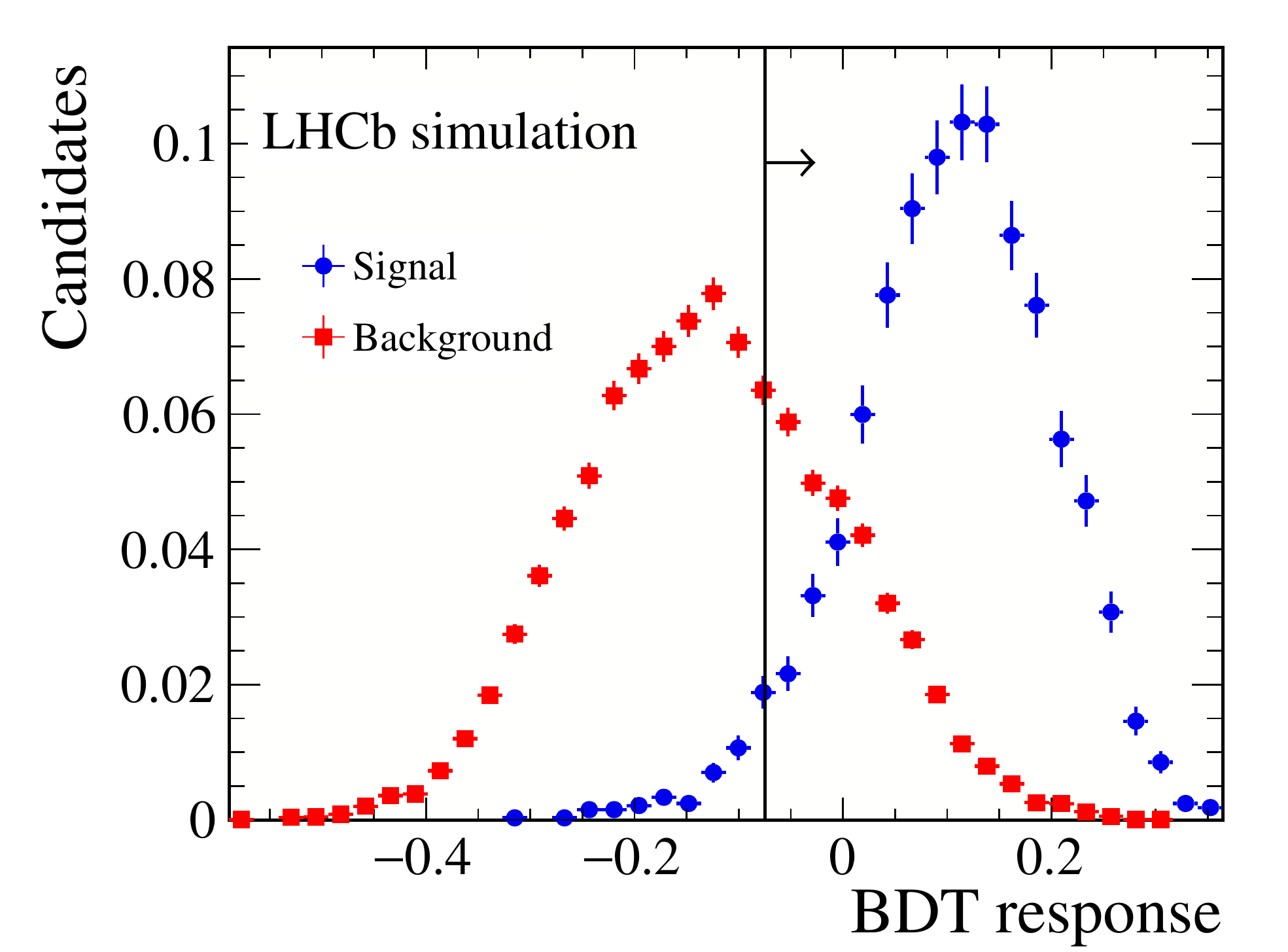}
  \caption{Distribution of the BDT response on the signal and background simulated samples.}
  \label{fig:bdtoutput}
\end{figure}

To suppress these background sources, LHCb has three very good tools: the $3\pi$ dynamics;
\begin{itemize}
  \item the isolation criteria against charged tracks and neutral energy deposits;
  \item partial reconstruction techniques in the signal and background hypotheses.
\end{itemize}
All these variables are used to train a boosted decision tree (BDT),
where $B^0\rightarrow D^{*-}\tau^+\nu_\tau$ simulation is used as signal, and double charm simulated decays,$ B\rightarrow D^* D X$, as background. The output of the BDT is shown in Fig.~\ref{fig:bdtoutput}.

\subsection{A model for $D_s^+\to 3\pi X$ decays}
\begin{figure}[ht]
  \centering
  \includegraphics[scale=0.7]{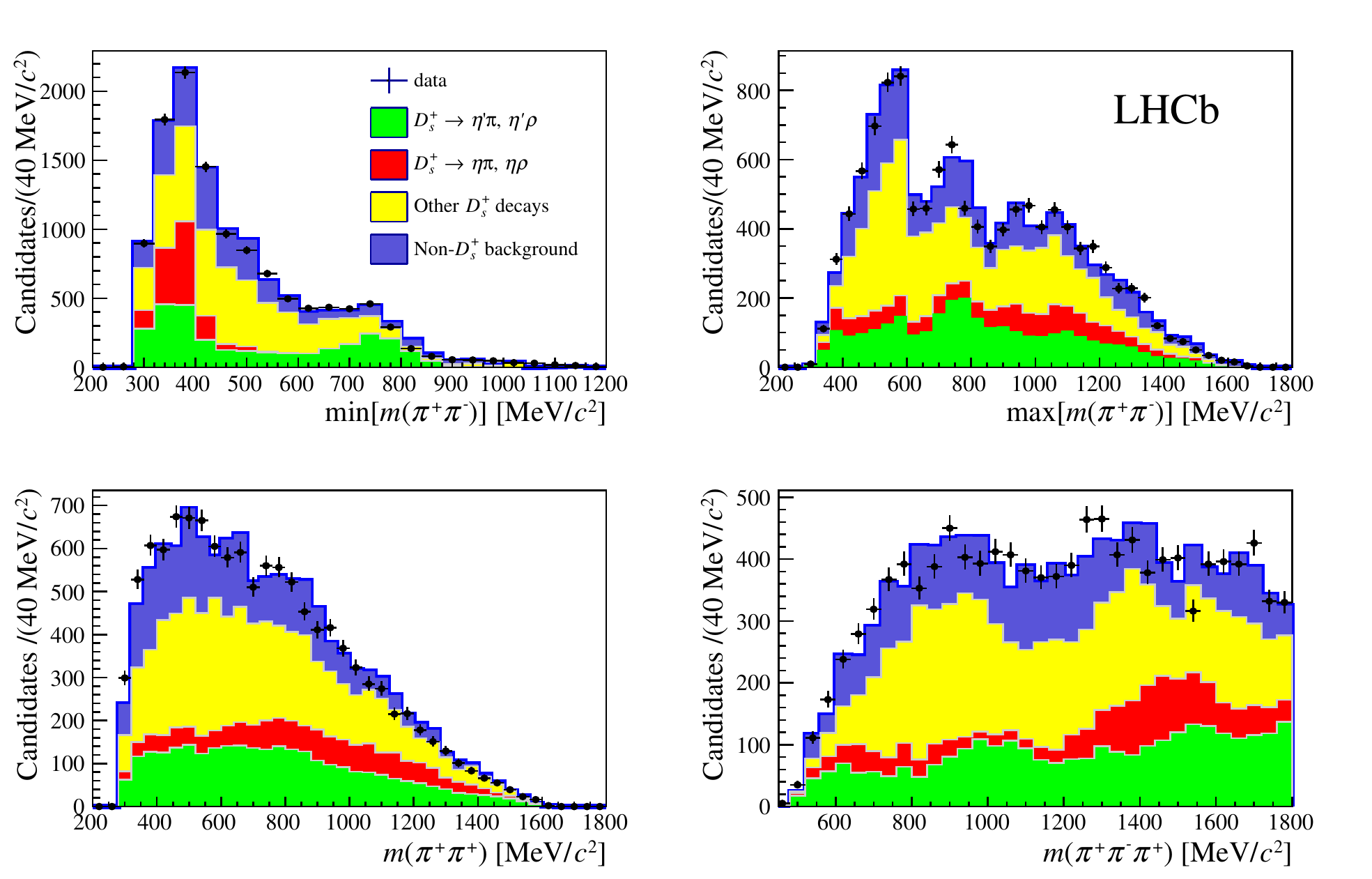}
  \caption{$D^+_s$ model fit on data. The different contributions are shown.}
  \label{fig:Dsmodel}
\end{figure}

The dominant double-charm background process $B \rightarrow D^{*-} D_s^+(X)$ is reduced by taking into account the resonant structure of the $3\pi$ system.
The $\tau$ lepton decays to $3\pi\nu_{\tau}$ final states predominantly through the
$a_1(1260)^+\rightarrow \rho^0 \pi^+$ decay. By contrast, the $D_s^+$ meson decays to $3\pi X$ final states predominantly through the $\eta$ and $\eta'$ resonances.
For this reason, the $D^+_s \rightarrow 3\pi X$ decay model has been determined directly from data, using a sample formed by events with low value of the BDT output, which is enriched in such decays. In order to retrieve the sub-decays relative fractions, a simultaneous fit to the minimum and maximum of $\pi^+\pi^-$ mass distribution, the invariant mass of the 2 pions with the same charge the invariant mass of the $3\pi$ system has been performed. The fit result is shown in Fig.~\ref{fig:Dsmodel}.
The model contains the following components:
\begin{itemize}
  \item $D^+_s$ decays where at least 1 pion comes from the decay of a $\eta$ or $\eta'$ meson : $\eta\pi$, $\eta\rho$, $\eta'\pi$ and $\eta'\rho$;
  \item $D^+_s$ decays where at least 1 pion is from an IS (Intermediate Resonance) other than $\eta$ and $\eta'$: IS$\pi$, IS$\rho$ (IS could be $\omega$, $\phi$);
  \item $D^+_s$ decays where none of the 3 pions comes from an IS, subdivided in: $K^0 3\pi$, $\eta3\pi$, $\eta' 3\pi$, $\omega3\pi$, $\phi3\pi$, $3\pi$ non resonant final state.
\end{itemize}
The results obtained by the fit are then used to construct the $D^+_s$ template used in the final fit.

\section{Signal extraction}
\subsection{Signal reconstruction}
\begin{figure}[H]
  \centering
  \includegraphics[scale=0.5]{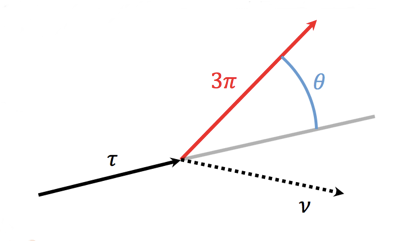}
  \caption{Schematic view of $\theta$, the angle between $\tau$ and $3\pi$ direction.}
  \label{fig:theta}
\end{figure}

Due to the presence of the neutrino in the final state of the signal decay, the $\tau$ momentum can be inferred, up to a two-fold ambiguity, by exploiting the flight direction of the $\tau$. In this way it is possible to compute the $\tau$ momentum with the following equation:

\begin{equation}
    |\vec{p}_{\tau}| = \frac{(m_{3\pi}^2+m_{\tau}^2)|\vec{p}_{3\pi}|\cos{\theta}
    \pm E_{3\pi}\sqrt{(m_{\tau}^2-m_{3\pi}^2)^2-4
    m_{\tau}^2|\vec{p}_{3\pi}|^2\sin^2\theta}}{2(E_{3\pi}^2-|\vec{p}_{3\pi}|^2\cos^2\theta)}
  \label{eq:ptau}
\end{equation}
where $\theta$ is the angle between $\tau$ and $3\pi$ direction as shown in Fig.~\ref{fig:theta}.

The ambiguity can be resolved by choosing the maximum value for the opening angle  between the three charged pions system and the direction of the $\tau$ lepton, as in the following equation:

\begin{equation}
  \theta_{max} = \arcsin \left(\frac{m_{\tau}^2-m_{3\pi}^2}{2m_{\tau}|\vec{p}_{3\pi}|}\right)
\end{equation}
This introduces a negligible bias in the signal reconstruction.

\subsection{Fit model and results}
An extended maximum likelihood 3-dimensional fit, using templates has been performed. The fitted variables are:
\begin{itemize}
  \item $q^2$, the invariant mass squared of the $\tau-\nu$ system
  \item $\tau$ decay time;
  \item The output of the BDT.
\end{itemize}

The fit model used to extract the signal yield consists of 5 main categories:
\begin{itemize}
  \item Signal described by the sum of two $\tau$ decays, $\tau^+ \rightarrow 3\pi\bar{\nu}_\tau$ and $\tau^+ \rightarrow 3\pi\pi^0 \bar{\nu}_\tau$;
  \item $B \rightarrow D^{**}\tau^+\nu_{\tau}$ decays;
  \item $B \rightarrow D^*D^+_s X$, $B \rightarrow D^*D^0 X$ and $ B\rightarrow D^*D^+X$ decays;
  \item $B\rightarrow D^{*-}3 \pi X$ decays;
  \item Combinatorial background.
\end{itemize}

\begin{figure}[H]
  \begin{center}
    \includegraphics[width=0.55\textwidth]{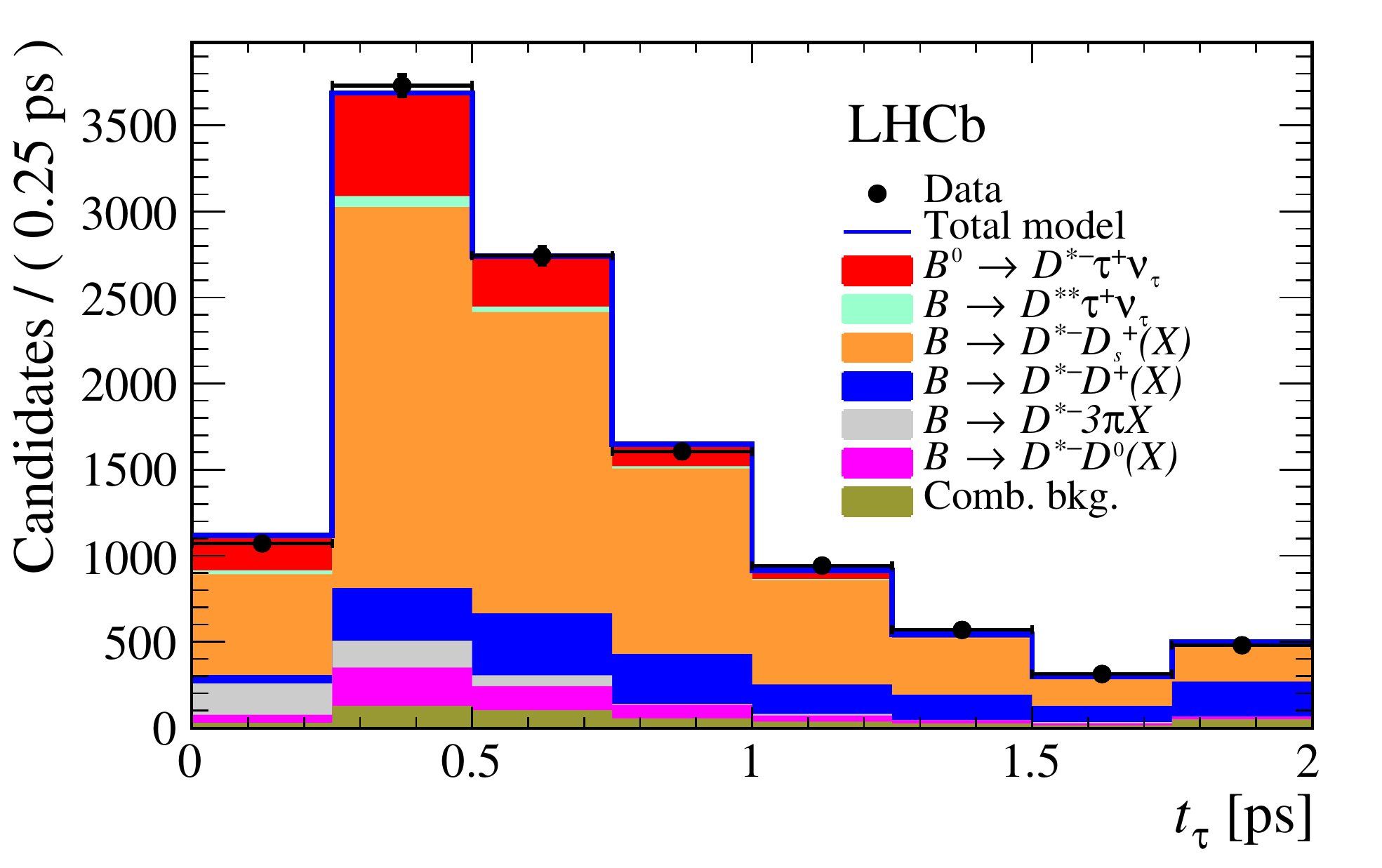}
    \includegraphics[width=0.55\textwidth]{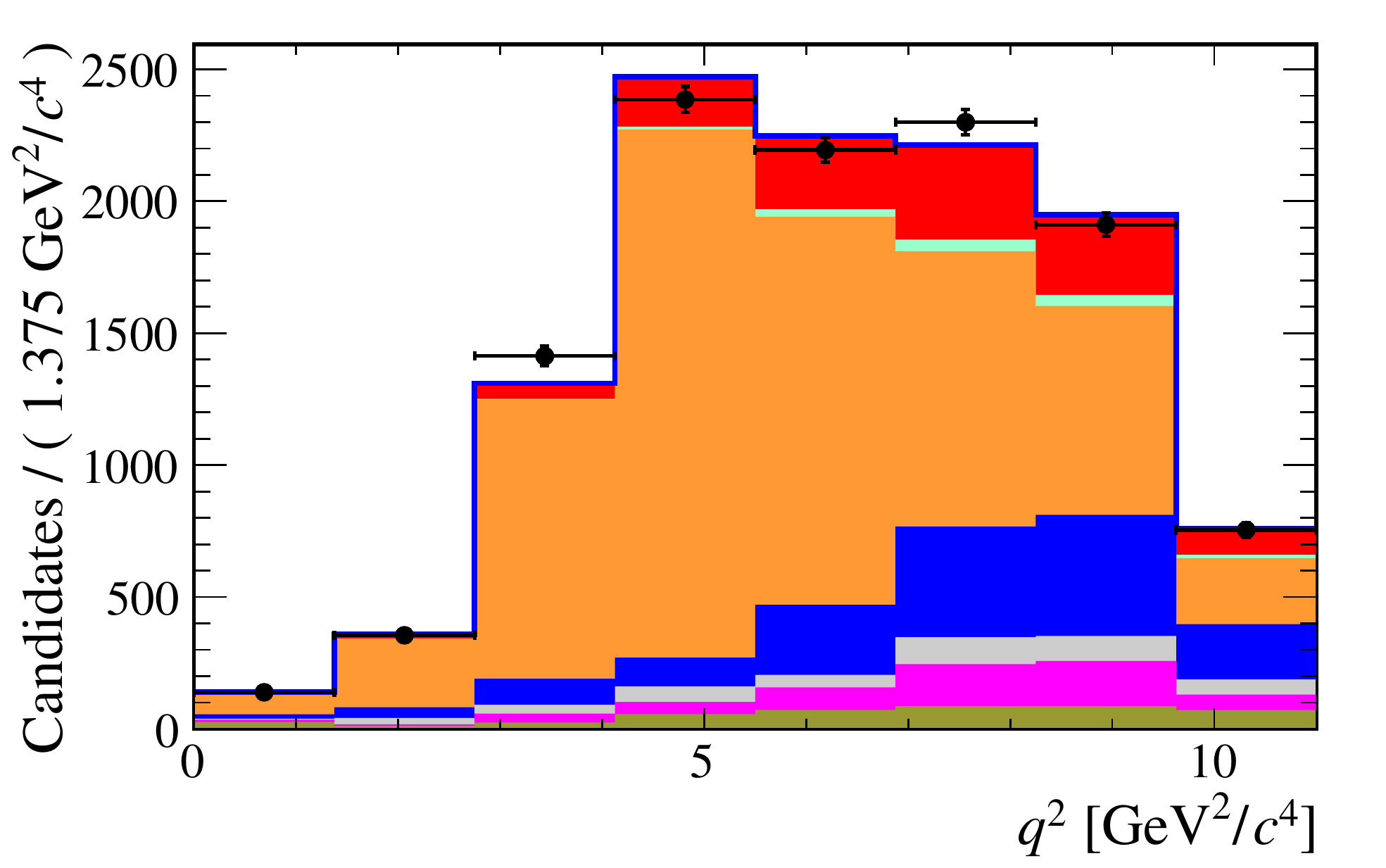}
    \includegraphics[width=0.55\textwidth]{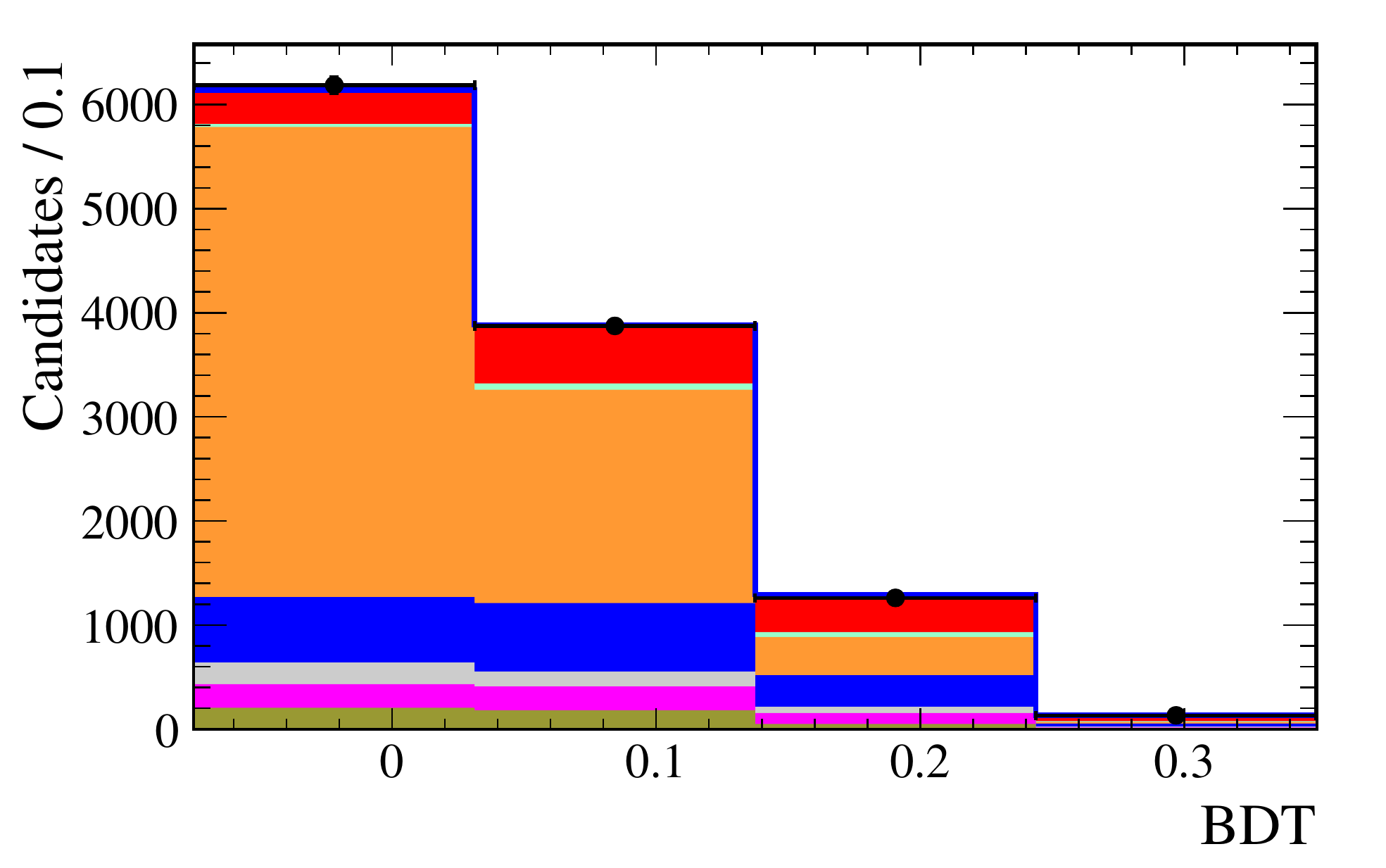}
  \end{center}
  \caption{
    \small 
  Projections of the three-dimensional fit on the (top) $3\pi$ decay time,
(middle) $q^2$ and (bottom) BDT output  distributions. The various components
are described in the legend.
 }
  \label{fig:fit_results}
\end{figure}

\begin{figure}[H]
  \begin{center}
    \includegraphics[width=0.9\textwidth]{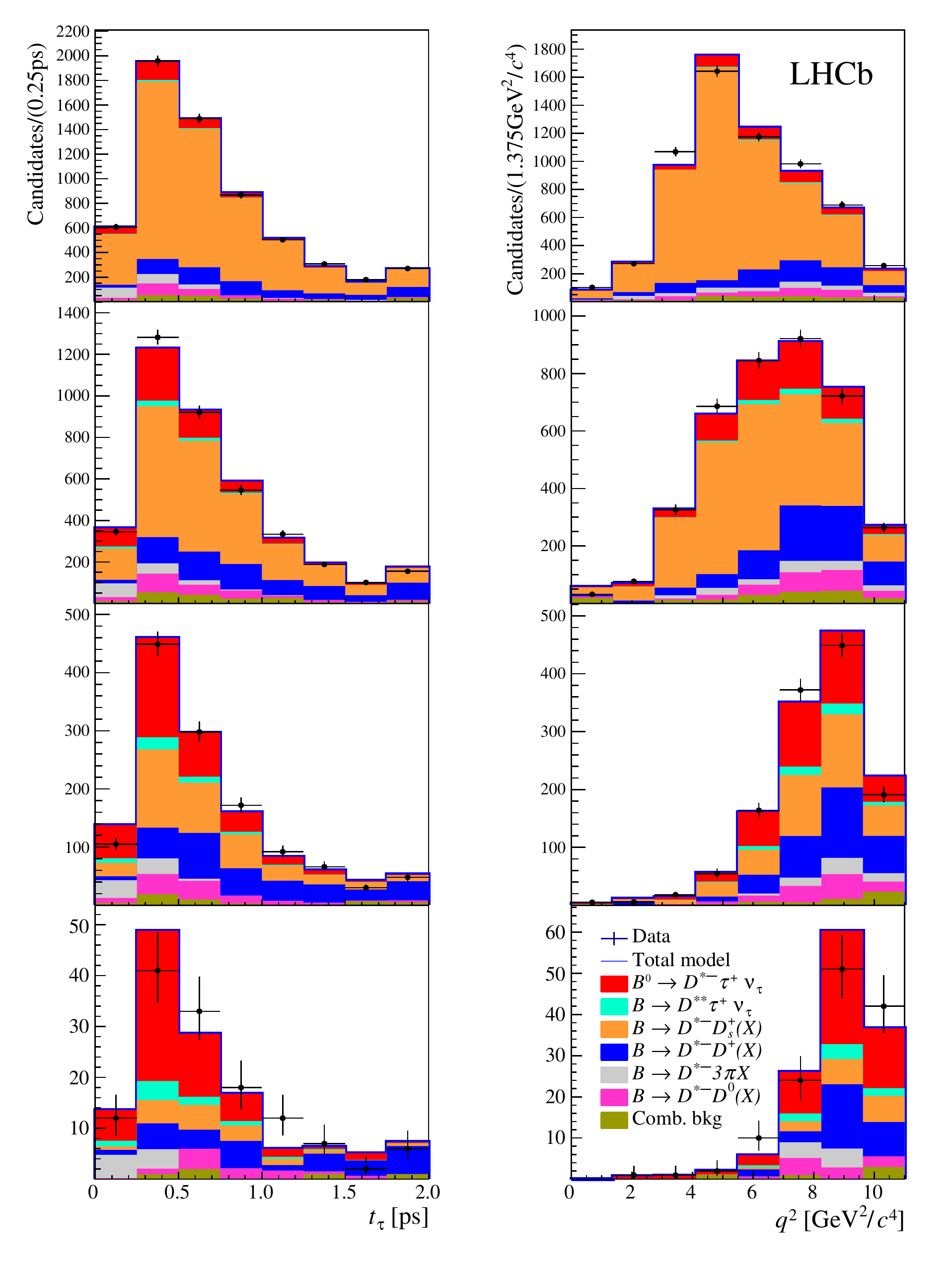}
  \end{center}
  \caption{
    \small 
Fit projections on the (left) $3\pi$ decay time  and (right) $q^2$ distributions for
the four different BDT bins shown in Fig.~\ref{fig:fit_results} (bottom). The highest BDT bin corresponds to the
two lower figures.
 }
  \label{fig:fit_bdtbins}
\end{figure}

The results of the three-dimensional fit are shown in Fig.~\ref{fig:fit_results}. A signal yield of $N_{sig} = 1273 \pm 85$ is found, where a $3\%$ correction due to fit bias has been applied.
Figure~\ref{fig:fit_bdtbins} shows the results on the fit in bins of the BDT output.


\section{Control samples}
\begin{figure}[H]
  \begin{center}
    \includegraphics[width=0.495\textwidth]{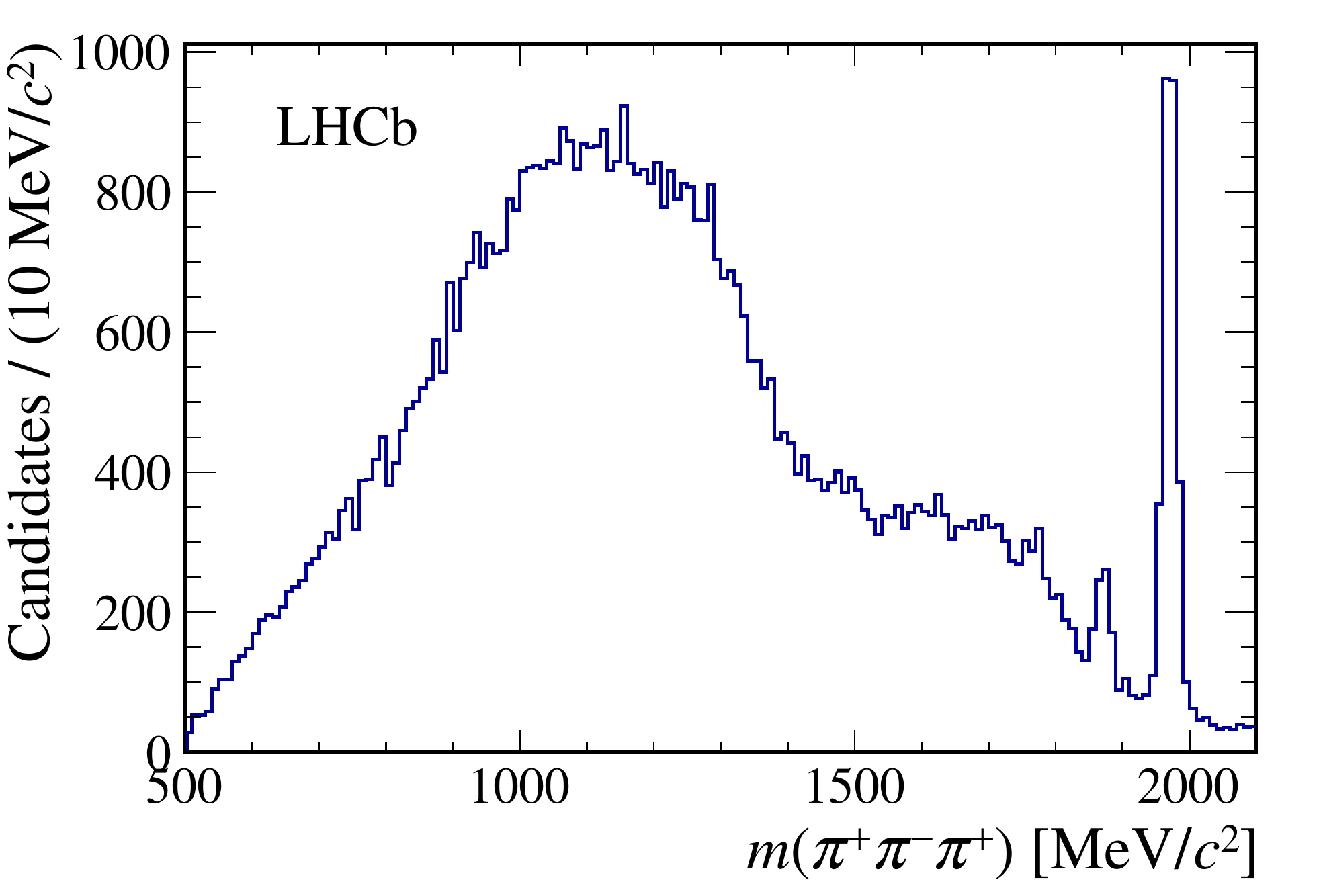}
    \includegraphics[width=0.495\textwidth]{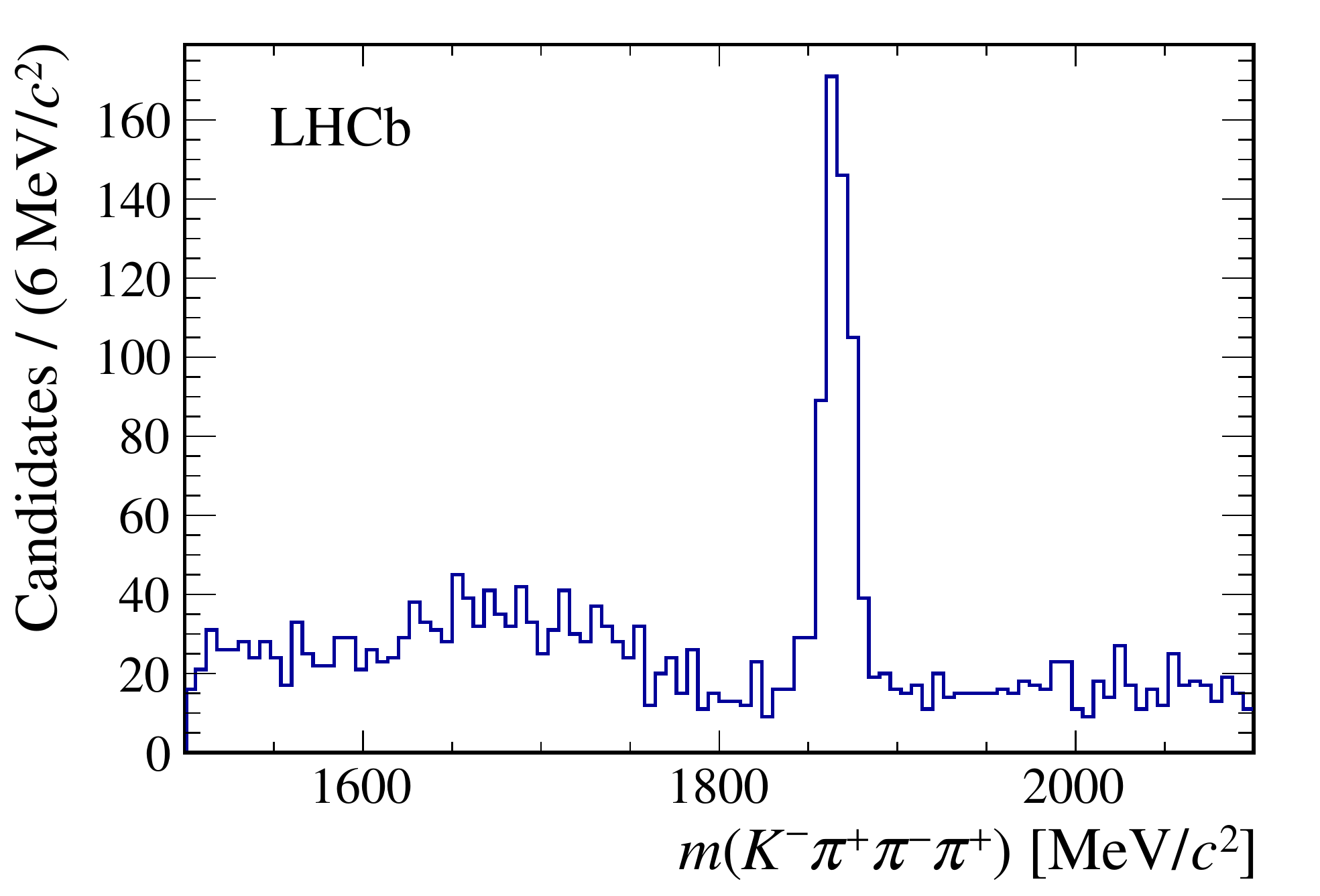}
    \includegraphics[width=0.495\textwidth]{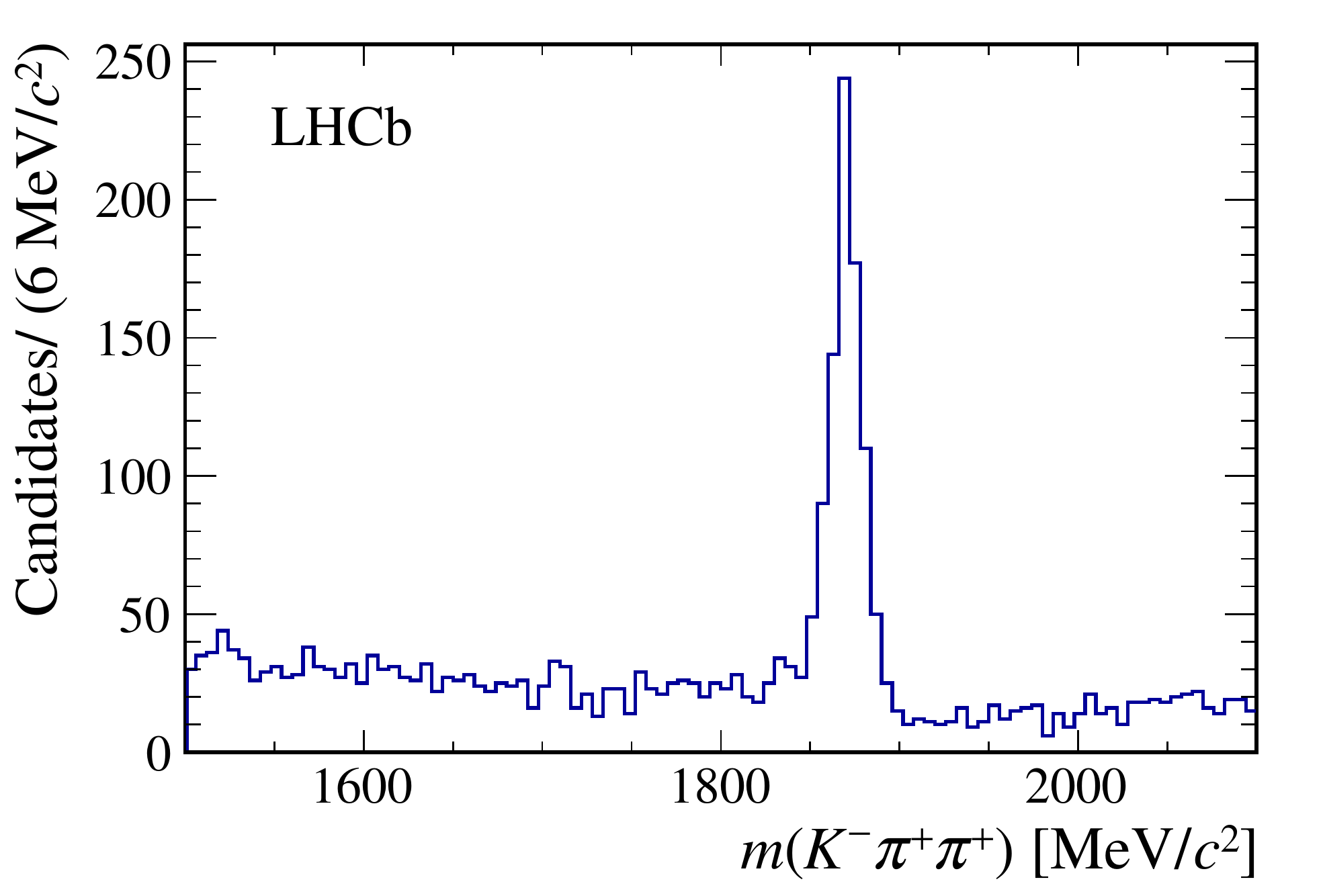}
  \end{center}
  \caption{
    \small 
      Data control samples for $D_s^+\rightarrow \pi^+ \pi^- \pi^+$ decays (top left), the $D^0\rightarrow K^- \pi^+ \pi^- \pi^+$ decay (top right) obtained requiring an extra particle in a cone around the $3\pi$ flight direction, and the $D^+ \rightarrow K^- \pi^+ \pi^+$ decays (bottom) obtained requiring the opposite charged pion to identified as a kaon.
    }
  \label{fig:control_samples}
\end{figure}

The fit that determines the signal yield uses templates that are taken from the simulation. It is therefore important to verify the agreement between data and simulation for background processes. Data control samples are used wherever is possible for this purpose. The relative contributions of double-charm backgrounds and their $q^2$ distribution from simulation are validated, and corrected where appropriate, by using data control samples enriched in such processes. In this way it is also possible to study inclusive decays of $D^0$, $D^+$ and $D_s^+$ mesons to $3\pi$. In Fig.~\ref{fig:control_samples} are shown the data control samples for $D_s^+\rightarrow \pi^+ \pi^- \pi^+$ decays (top left), the $D^0\rightarrow K^- \pi^+ \pi^- \pi^+$ decay (top right) obtained requiring an extra particle in a cone around the $3\pi$ flight direction, and the $D^+ \rightarrow K^- \pi^+ \pi^+$ decays (bottom) obtained requiring the opposite charged pion to identified as a kaon.

\section{Normalization mode}
\begin{figure}[H]
\centering
 \includegraphics[width=0.6\textwidth]{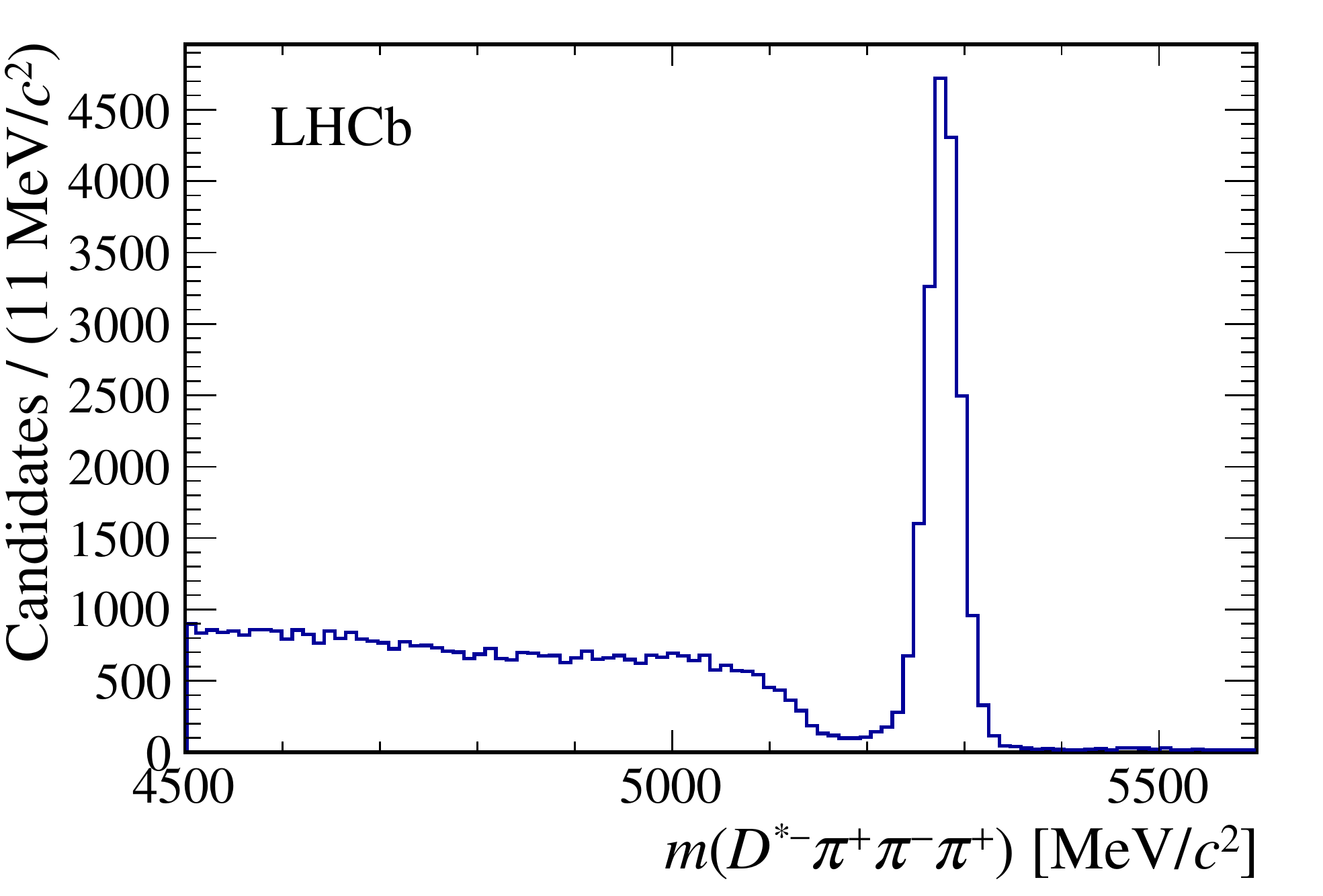}
\caption{Distribution of the $D^{*-} 3\pi$ mass for candidates passing the selection.}
\label{fig:normalization}
\end{figure}

The normalization channel has to be as similar as possible to the signal channel in order to cancel all systematics linked to trigger selection, particle identification and selection cuts.
The two decays differ by the presence of a softer pion and $D^*$ in the signal decay due to the two extra neutrinos, and the kinematics of the $3\pi$ system. This will give a residual effect on the efficiency ratio. Figure \ref{fig:normalization} shows the $D^{*-}3\pi$ invariant mass after the selection of the normalization sample. A clear $B^0$ peak signal peak is seen. In order to extract the normalization yield, a fit is performed in the 5150-5400 MeV/$c^2$ region. The signal is described by the sum of a Gaussian and a Cyrstal Ball function. An exponential function is used to describe the background. The yield obtained is $N_{norm} = 17808 \pm 143$.

\section{Systematic uncertainties}

Systematic uncertainties on $K(D^{*-})$ are reported in Table~\ref{tab:systematics}. The main systematic uncertainties are due to the limited size of the simulated samples that contributes to the efficiencies and to the determination of the template shapes for signal and background. The presence of empty bins in the templates used in the fit introduces a positive bias in the determination of the signal yield that is corrected for.
In the uncertainty computation, another contribution is given by the knowledge of the $B\rightarrow D^{*-}(D_s,D^0,D^+)X$ and $B \rightarrow D^{*-}3\pi X$ background shapes. When computing $R(D^{*-})$, an additional systematic of $4.5\%$ due to the knowledge of the external branching fration must be taken into account. The size of the total systematics is above the statistical precision but there is room for improvements, i.e., with more data and simulated samples, and a better knowledge of the external branching franctions using also results from other experiments.

\begin{table}
  \centering
  \caption{Relative systematic uncertainties on ${\cal{K}}(D^{*-})$.}
  \label{tab:systematics}
    \begin{tabular}{l|c}
      \hline
      Source                          & $\delta K(D^{*-}) /
                                        K(D^{*-}) [\%]$ \\
      \hline
      Simulated sample size     & 4.7 \\
      Empty bins in templates                            & $1.3$ \\
      Signal decay model               &  1.8 \\
      $D^{**}\tau^+\nu_{\tau}$ and $D_s^{**}\tau^+\nu_{\tau}$ feeddowns & $2.7$ \\
      $D_s^+ \to 3\pi X$ decay model         & $2.5$ \\
      $B\rightarrow D^{*-} D_s^+ X$, $B\rightarrow D^{*-} D^+ X$, $B\rightarrow D^{*-} D^0 X$ backgrounds & 3.9 \\
      Combinatorial background              & $0.7$ \\
      $B\rightarrow D^{*-} 3\pi X$ background & $2.8$ \\
      Efficiency ratio                  & 3.9 \\
        \hline
      Total uncertainty                  & $8.9$ \\
      \hline
    \end{tabular}
\end{table}

\section{Conclusions}

The first measurement of $R(D^{*-})$ with three pions $\tau$ decays has been performed, using a technique that is complementary to all previous measurements of this quantity and offer the possibility to study other b-hadron decay modes in a similar way.
The ratio of branching fractions between the $B^0 \rightarrow D^{*-} \tau^+ \nu_\tau$ and the $B^0\rightarrow D^{*-}3\pi$ decays is measured to be
\begin{align*}
  K(D^{*-}) = 1.93 \pm 0.13 \textnormal{ (stat)} \pm 0.17\textnormal{ (syst)},
\end{align*}
where the first uncertainty is statistical and the second systematic. The first determination of $R(D^{*-})$ performed by using 3-pion $\tau$ decays is obtained by using the measured branching fractions of
$\mathcal{B}(B^0\rightarrow D^{*-} \mu^+\nu_\mu) = (4.88 \pm 0.10) \times 10^{-2}$ and $\mathcal{B}(B^0 \rightarrow D^{*-}3\pi) = (7.21 \pm 0.29)\times 10^{-3}$, with the result: $R(D^{*-}) = 0.285 \pm 0.019 \textnormal{ (stat) } \pm 0.025 \textnormal{ (syst) } \pm 0.013\textnormal{ (ext)}$, is one of the most precise single measurement performed so far. It is 1.0 standard deviation higher than the SM prediction $0.252 \pm 0.003$, and consistent with previous determinations.

\bibliographystyle{unsrt}
\bibliography{article}

%
%
%
%
%
%
%
%
%
%
%
%
%
%

\def\Discussion{
\setlength{\parskip}{0.3cm}\setlength{\parindent}{0.0cm}
     \bigskip\bigskip      {\Large {\bf Discussion}} \bigskip}
\def\speaker#1{{\bf #1:}\ }
\def\endDiscussion{}

\end{document}